\documentclass[runningheads]{llncs}
\usepackage{graphicx}
\usepackage{url}
\begin{document}
\title{Deep Learning for Energy Estimation and Particle Identification in Gamma-ray Astronomy\thanks{Supported by the Russian Science Foundation, project 18-41-06003.}}
\titlerunning{Deep Learning in Gamma-Ray Astronomy}
\author{
Evgeny Postnikov\inst{1}\orcidID{0000-0002-3829-0379} \and
Alexander Kryukov\inst{1} \orcidID{0000-0002-1624-6131} \and
Stanislav Polyakov\inst{1} \orcidID{0000-0002-8429-8478} \and
Dmitry Zhurov\inst{2,3} \orcidID{0000-0002-1596-8829}
}
\authorrunning{E. Postnikov et al.}
\institute{Lomonosov Moscow State University, Skobeltsyn Institute of Nuclear Physics (SINP MSU), 1(2), Leninskie gory, GSP-1, Moscow 119991, Russian Federation\\
\email{evgeny.post@gmail.com, kryukov@theory.sinp.msu.ru, s.p.polyakov@gmail.com}\\
\url{http://www.sinp.msu.ru/en} 
\and
%Irkutsk State University, 1, Karl Marx St. 664003 Irkutsk, Russian Federation
%\email{sidney28@ya.ru}\\
%\url{http://isu.ru/en}
%\and
Institute of Applied Physics of Irkutsk State University, 20, Gagarin bulvard 664003 Irkutsk, Russian Federation\\
\email{sidney28@ya.ru}\\
\url{http://api.isu.ru}
\and
Irkutsk National Research Technical University, 83, Lermontova str. 664074 Irkutsk, Russian Federation\\
%\email{sidney28@ya.ru}\\
\url{http://www.istu.edu/eng}} %\and
%\email{sidney28@ya.ru}}
%
\maketitle              % typeset the header of the contribution
\begin{abstract}

Deep learning techniques, namely convolutional neural networks (CNN), have previously been adapted to select gamma-ray events in the TAIGA experiment, having achieved a good quality of selection as compared with the conventional Hillas approach. Another important task for the TAIGA data analysis was also solved with CNN: gamma-ray energy estimation showed some improvement in comparison with the conventional method based on the Hillas analysis. Furthermore, our software was completely redeveloped for the graphics processing unit (GPU), which led to significantly faster calculations in both of these tasks. All the results have been obtained with the simulated data of TAIGA Monte Carlo software; their experimental confirmation is envisaged for the near future.

\end{abstract}

\section{Introduction}

\subsection{Gamma-ray astronomy}

Gamma-ray detection is very important for observing the Universe as gamma-rays are particles without electric charge and are unaffected by a magnetic field. Detected gamma-rays can therefore be extrapolated back to their origin. For that reason, they are currently the best "messengers" of physical processes from the relativistic Universe.

With specially designed telescopes, gamma-rays can be detected on Earth (ground-based gamma-ray astronomy) at very high energies. These instruments are called Imaging Air Cherenkov Telescopes (IACTs) \cite{1}.  Gamma-rays are observed on the ground optically via the Cherenkov light emitted by extensive showers of secondary particles in the air when a very-high-energy gamma-ray strikes the atmosphere.  

However, very-high-energy gamma-rays contribute only a minuscule fraction to the flux of electrically charged cosmic rays (below one per million \cite{2}). This circumstance makes it necessary to learn to distinguish gamma-rays against charged cosmic rays, mostly protons, on the basis of the images they produce in the telescope camera. 

\subsection{Data Life Cycle project in Astroparticle Physics}
The Russian-German Initiative of a Data Life Cycle in Astroparticle Physics (also referred to as Astroparticle.online) \cite{3,4} aims to develop an open science system for collecting, storing, and analyzing astroparticle physics data including gamma-ray astronomy data. Currently it works with the TAIGA \cite{5} and KASCADE \cite{6} experiments and invites astrophysical experiments to participate.

In this work, two important problems of gamma-ray astronomy data analysis are solved within the framework of deep learning approach (convolutional neural networks). These are the background rejection problem (removal of cosmic ray background events), and the gamma-ray energy estimation problem, in imaging air Cherenkov telescopes. The data to solve the both problems were simulated using the complete Monte Carlo software for the TAIGA-IACT installation \cite{7}.

\subsection{Convolutional Neural Networks (CNNs)} 
CNNs are well adapted to classify images; that is why they were also chosen for all deep learning applications to the IACT technique \cite{8,9,10}. Their advantage is a fully automatic algorithm, including automatic extraction of image features instead of a set of empirical parameters  (`Hillas parameters' \cite{11}). CNNs are implemented in various free software packages, including PyTorch \cite{12} and TensorFlow \cite{13}. In contrast to the camera with square pixels \cite{8}, a shape and arrangement of pixels of the TAIGA-IACT camera is hexagonal, and this geometrical feature has not been fully taken into account yet.

\section{Data simulations}

Data simulations were performed to obtain datasets with the response of a real IACT telescope for two classes of particles to be identified: gamma-rays and background particles (protons). The development of the shower of secondary particles in the atmosphere was simulated with the CORSIKA package \cite{14}. The response of the IACT system was simulated using the OPTICA-TAIGA software developed at JINR, Dubna \cite{7}. It describes the real TAIGA-IACT setup configuration: 29 constituent mirrors with an area of about 8.5 m$^2$ and a focal length of 4.75 m, and the 560-pixel camera located at the focus. Each pixel is a photomultiplier (PMT) collecting light from the mirrors. 

The telescopic image was formed using a dedicated software developed at SINP MSU taking into account the night sky background fluctuations, PMT characteristics, and triggering and readout procedures of the data acquisition system.

\section{Image cleaning}

Image cleaning is a conventional procedure to remove images and image parts produced by the night sky background fluctuations but not by a shower of secondary particles. The conventional procedure is two-parametric: it excludes from subsequent analysis all image pixels except the “core pixels”, i.e. those with the amplitude above a “core threshold” and at least one neighbour pixel above a “neighbour threshold”, and the neighbour pixels themselves. If the image contains too few pixels after cleaning (for example, 2 or less), the entire image is excluded from the analysis.

Deep learning algorithms were trained on images both without and with cleaning. For the reference technique, a test sample was first subjected to the image cleaning procedure in any case. No training sample was needed for the reference technique.

\section{Deep learning}

\subsection{Data sample} 

\begin{figure}
\begin{center}
 \begin{minipage}{1\textwidth}
  \includegraphics[width=14pc]{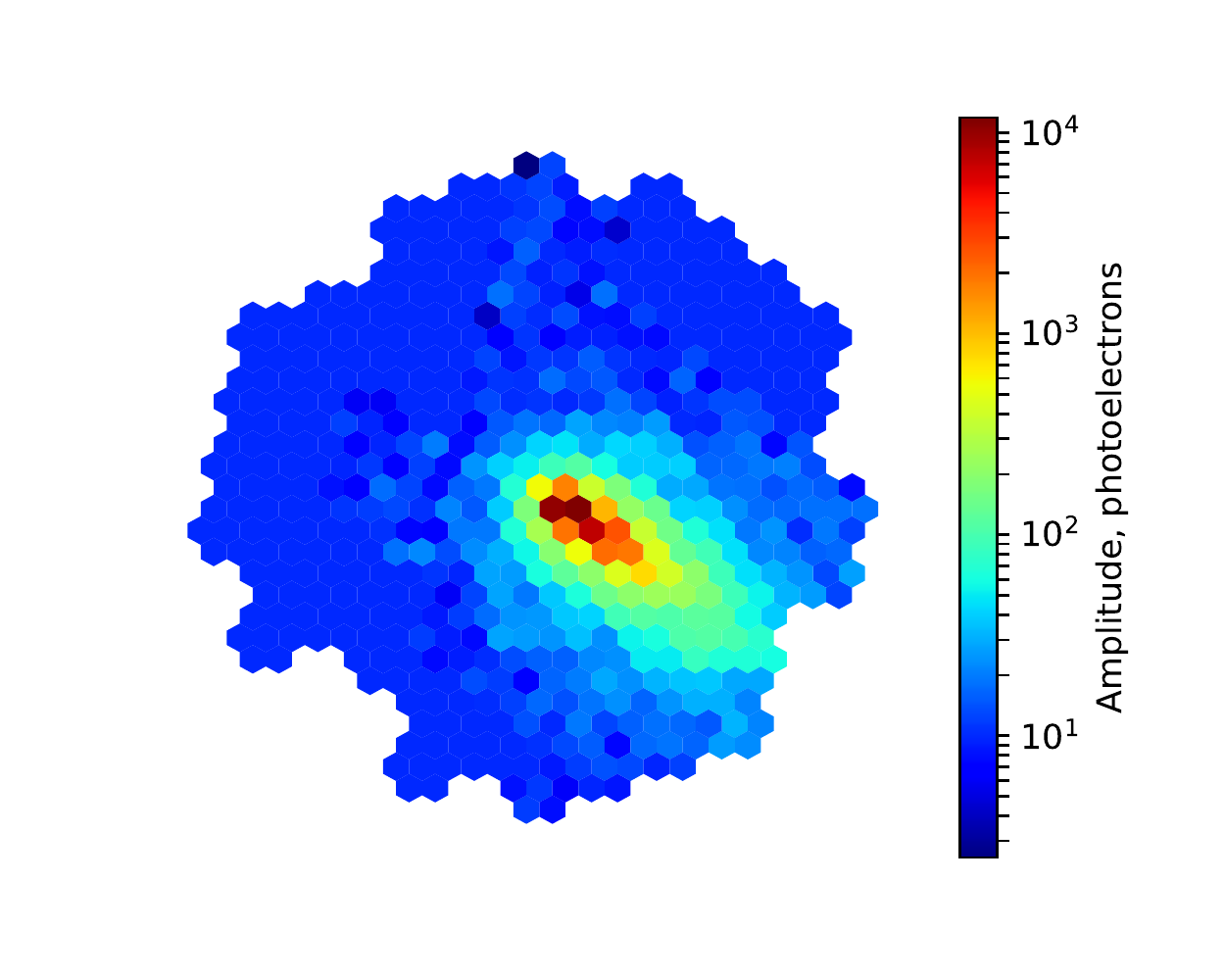}
  \includegraphics[width=14pc]{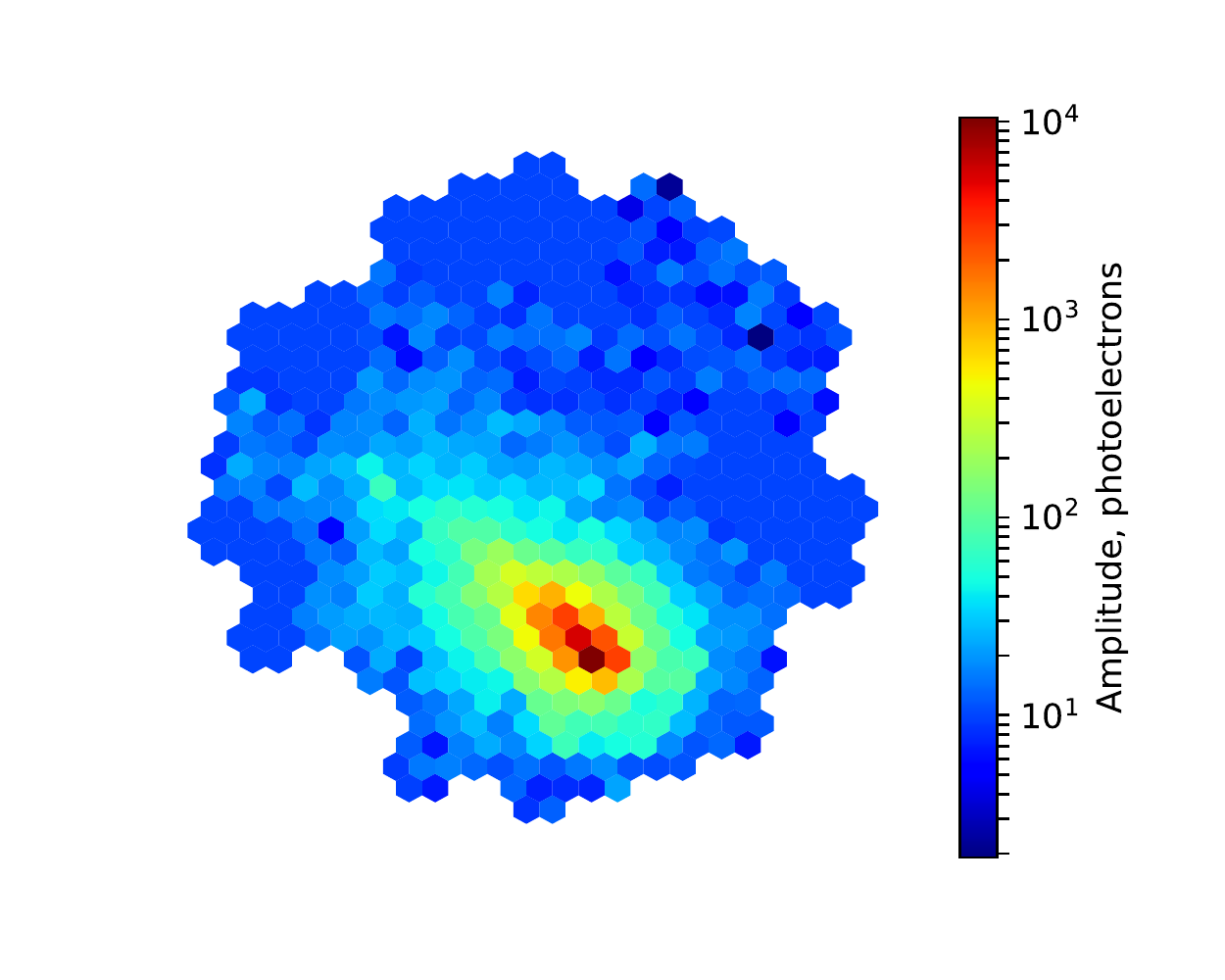}
  %\caption{second}
 \end{minipage}
\end{center}
\caption{Simulated image examples: generated by high-energy gamma-ray (left) and proton (right).} 
\label{image_ex}
\end{figure}

Training datasets for the CNN contained gamma-ray and proton images (Monte Carlo of TAIGA-IACT) for the task of background suppression, and only gamma-ray images for the energy estimation. Image examples are presented in Figure \ref{image_ex}.

The dataset consisted of 2.7$\times$10$^4$ simulated events after strong image cleaning (70\% training + 30\% test) for PyTorch, and of 5.6$\times$10$^4$ events after soft image cleaning (of which 60\% were used for training, 15\% for validation, and 25\% for testing) for TensorFlow. The images in the training dataset were rotated around the camera center by multiples of 60$^o$ thereby producing 6 times the initial sample size. Finally, the total number of events was about 2$\times$10$^5$ for training (with rotations), 0.8$\times$10$^4$ for validation, and 1.5$\times$10$^4$ for testing.

\subsection{CNN implementation}  \label{secImplem}

Seeing that convolutional operations are optimized for a square grid, the TAIGA-IACT hexagonal camera grid was needed to be represented in convenient form to fit the square one. For that purpose, a transformation to oblique coordinate system was applied to each image, so that each hexagonal image with 560 pixels was transformed to the 31x30 square grid.  These square grid images were fed to the input layer of the CNN. 

For the background suppression, test datasets of gamma-ray and proton images in random proportion (blind analysis) were classified by each of the packages: TensorFlow and PyTorch. 

The energy was either directly predicted as a scalar parameter by the CNN, or the ratio of the energy to the total sum of the amplitudes in the image was predicted and then multiplied back by the value of the total sum to obtain the energy estimate. The reason for the second way to estimate the energy is that the above mentioned total sum of the amplitudes, referred to as `image size', is correlated with the energy for gamma-rays incident closer than $\approx$100~m from the telescope \cite{15}.  Therefore, image size can be in some way directly included in the estimation algorithm to account for this strong correlation at least for nearby gamma-rays. Beyond this `Cherenkov radius' of about 100--120 m, the Cherenkov light intensity varies rapidly with the distance from gamma-ray to the telescope, which may also lead to a substantial increase of the resulting uncertainty in the energy estimation.
      
Various networks with different parameters were tested to find the one maximizing background suppression and the one minimizing the relative energy error. 

\subsection{CNN architecture}

\begin{figure}
\includegraphics[width=\textwidth]{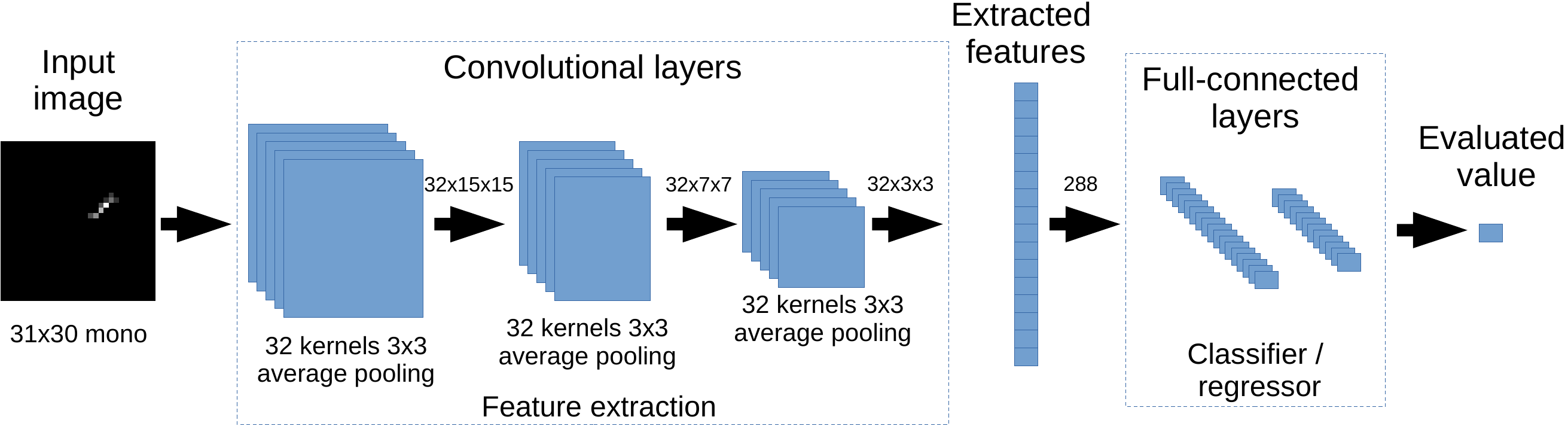}
\caption{Convolutional neural network for classification/regression. The network accepts square grid images in oblique coordinate system at the input of convolutional layers. Output of the convolutional layers (extracted features) is fed to the classifier/regressor (full-connected layers) that evaluate the output value.} \label{cnn}
\end{figure}

The first part of the convolutional neural network consists of convolutional layers (Figure \ref{cnn}). Each layer includes convolution with 32 kernels of 3x3 pixels and the ReLU activation function, average pooling layer with 3x3 pooling size and strides of 2 pixels. To avoid overfitting during the training, a dropout layer with a dropout rate of 10\% is added after each pooling layer. 

Output of the convolutional layer is fed to the full-connected layers of classifier or regressor. The full-connected layers consist of 32 neurons with the ReLU activation function in the first layer and 16 neurons in the second one. Dropout with a 50\% rate after each full-connected layer is used to avoid overfitting.

Sigmoid was set as the output neuron activation function for the classification task, whereas no activation function was set to the output neuron for the energy estimation. Adagrad optimizer with the learning rate set at 0.05 and the binary cross-entropy as the loss function were used for classification. The energy estimation was performed using Adam optimizer and the mean square error as the loss function. 

The early stop criterion was set to interrupt the training procedure when the loss function for the validation dataset shows no decrease for 30 epochs. The training lasted for 144 epochs (runtime $\sim$9 minutes). The computational graph was run on NVIDIA GPU Tesla P100.

Accuracy on the training and validation sample after training was 91.29\% and 90.02\% respectively. ROC AUC score (an area under the receiver operating characteristic curve \cite{16}) was 0.9712 for training and 0.9647 for validation.

\section{Results}

\subsection{Scalar quality criterion (Q-factor)}

As a quality criterion of particle identification, the selection quality factor $Q$ was estimated. This factor indicates an improvement of a significance of the statistical hypothesis that the events do not belong to the background in comparison with the significance before selection. For Poisson distribution (that is for a large number of events), the selection quality factor is:
\begin{equation}
Q=\epsilon_{nuclei}/\sqrt{\epsilon_{bckgr}},
\end{equation}
where $\epsilon_{nuclei}$ and $\epsilon_{bckgr}$ are relative numbers of selected events and background events after selection. For our task we consider protons as a background to select gamma-rays above this background. 

Quality factor (1) values obtained by the best CNN configuration among all the trained networks are assembled in Table \ref{tab1} together with the quality factor for the reference technique (simplest Hillas analysis \cite{17}).

CNN was accelerated on graphics processing unit (GPU), which led to a significantly faster calculation. Its implementation was approximately 6 times faster than an equivalent implementation on CPU and revealed no quality loss (the last column of Table \ref{tab1}).

%\paragraph{Sample Heading (Fourth Level)}
%The contribution should contain no more than four levels of headings. Table~\ref{tab1} gives a summary of all heading levels.

\begin{table}
\caption{Q-factor for gamma/proton identification.}\label{tab1}
\begin{tabular}{|l|l|l|l|l|}
\hline
Image cleaning&Reference&CNN(PyTorch)&CNN(TensorFlow)&CNN(TensorFlow GPU)\\
\hline
No&1.76&1.74&1.48&\\
Yes&1.70&2.55&2.91&2.86\\
%Title (centered) &  {\Large\bfseries Lecture Notes} & 14 point, bold\\
%1st-level heading &  {\large\bfseries 1 Introduction} & 12 point, bold\\
%2nd-level heading & {\bfseries 2.1 Printing Area} & 10 point, bold\\
%3rd-level heading & {\bfseries Run-in Heading in Bold.} Text follows & 10 point, bold\\
%4th-level heading & {\itshape Lowest Level Heading.} Text follows & 10 point, italic\\
\hline
\end{tabular}
\end{table}

\subsection{Q-factor variability}

\begin{figure}
\begin{center}
\includegraphics[width=0.8\textwidth]{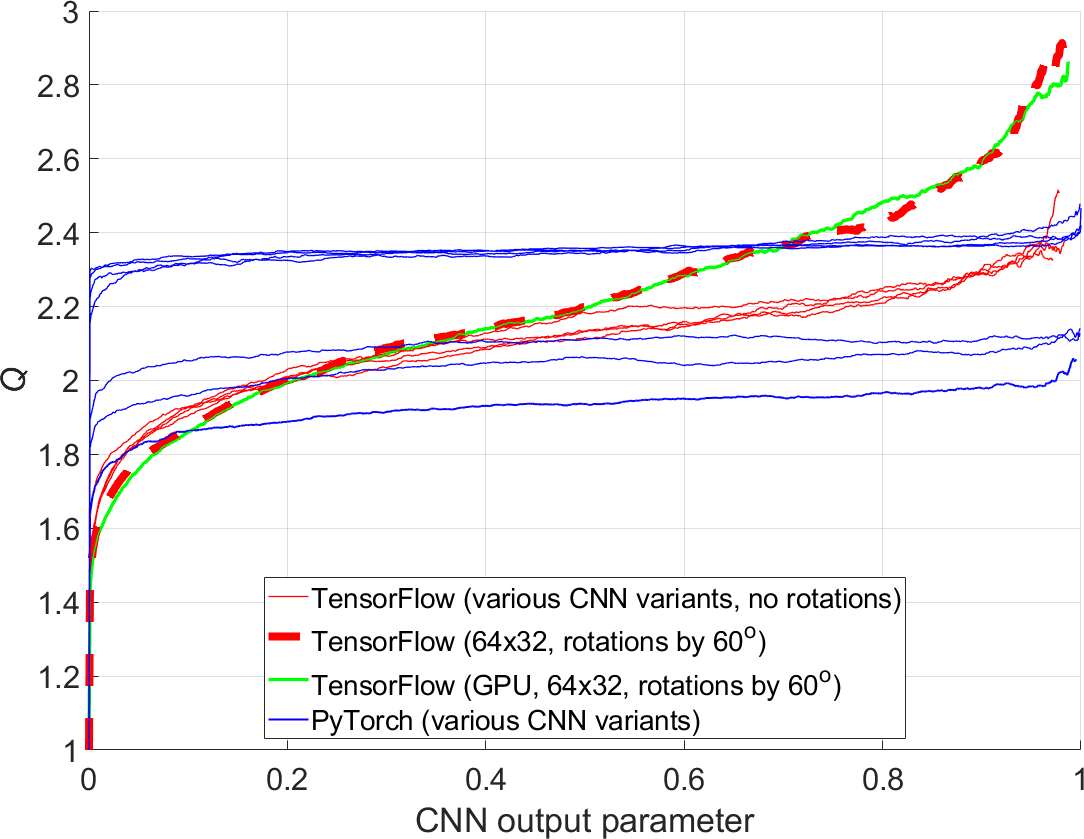}
\caption{Quality factor vs CNN output parameter (a scalar parameter between 0 and 1 characterizing image similarity to gamma-ray or proton).} \label{fig1}
\end{center}
\end{figure}

Comparison of different CNN versions for both software packages is illustrated in Figure \ref{fig1}. 

PyTorch had more stable results in a wide range of CNN output parameter values. However, significant improvement was obtained with the TensorFlow CNN version trained by a modified training sample, which contained both original simulated images and additional ones obtained by rotating images from the initial sample by the symmetry angles of hexagonal structure. Thus the modified training sample consisted of $\sim$2$\times10^5$ events instead of $\sim$3$\times10^4$. Therefore, the performance of different software packages was approximately the same, indicating that the training sample size was crucial for the identification quality. 
%\noindent Displayed equations are centered and set on a separate
%line.
%\begin{equation}
%x + y = z
%\end{equation}
%Please try to avoid rasterized images for line-art diagrams and
%schemas. Whenever possible, use vector graphics instead (see
%Fig.~\ref{fig1}).

\begin{figure}
\begin{center}
 \begin{minipage}{1\textwidth}
  \includegraphics[width=.88\textwidth]{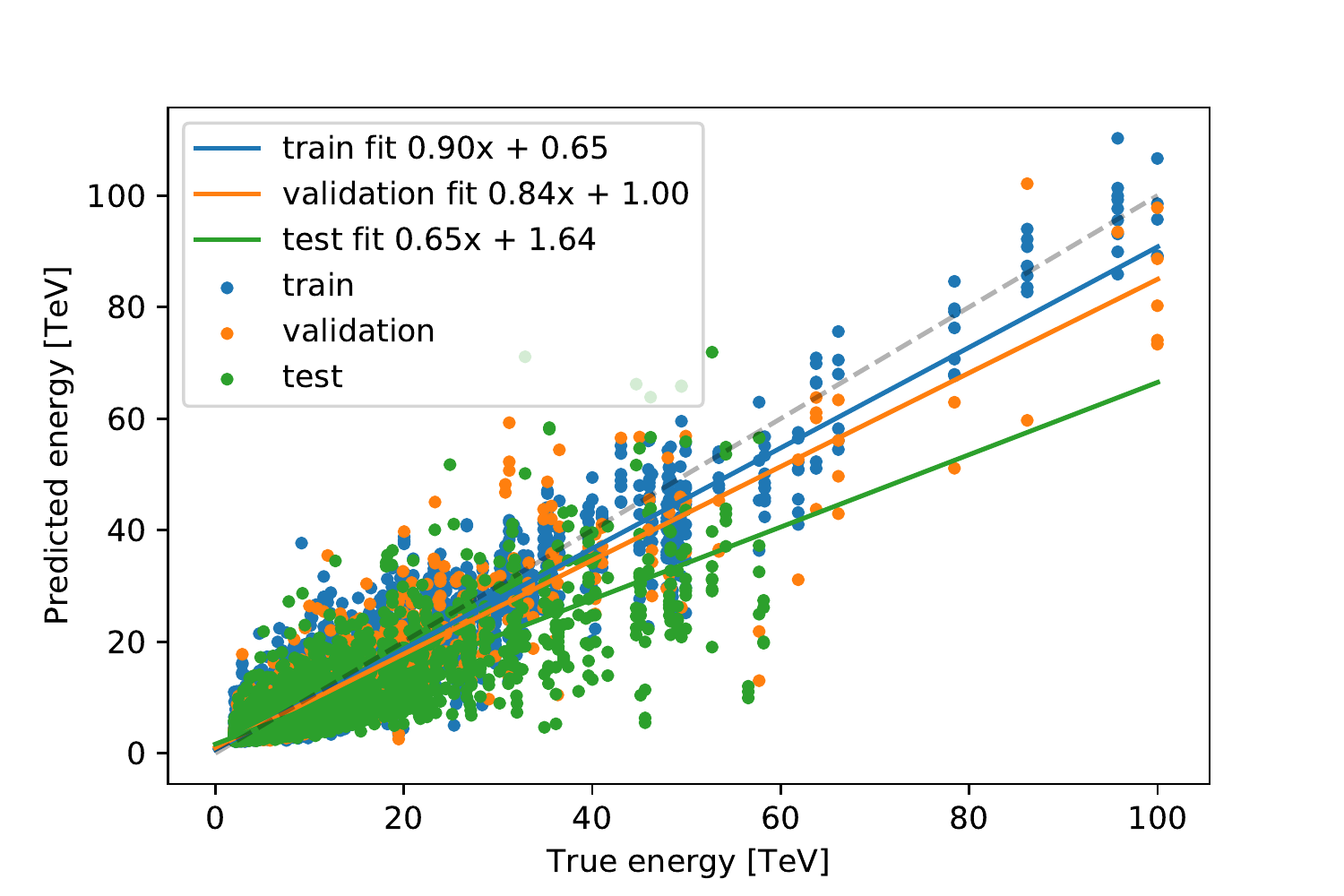}
  \includegraphics[width=.79\textwidth]{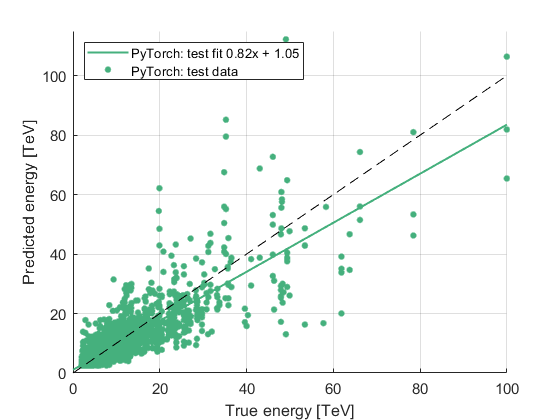}
 % \includegraphics[width=\textwidth]{rms_on_dist}
  %\caption{second}
 \end{minipage}
\end{center}
\caption{Predicted energy vs true energy: top pannel TensorFlow, bottom pannel PyTorch (dashed lines are the `ideal case' y=x).} 
\label{reg_im}
\end{figure}

\begin{figure}
\begin{center}
%\begin{left}
 %\begin{minipage}{1\textwidth}
%\includegraphics[width=\textwidth]{QCNNvsProb_GPU3crop.png}
%  \includegraphics[width=18pc]{delta_E_hist}
%  \includegraphics[width=11pc]{delta_E_hist_PTcrop}
  \includegraphics[width=0.8\textwidth]{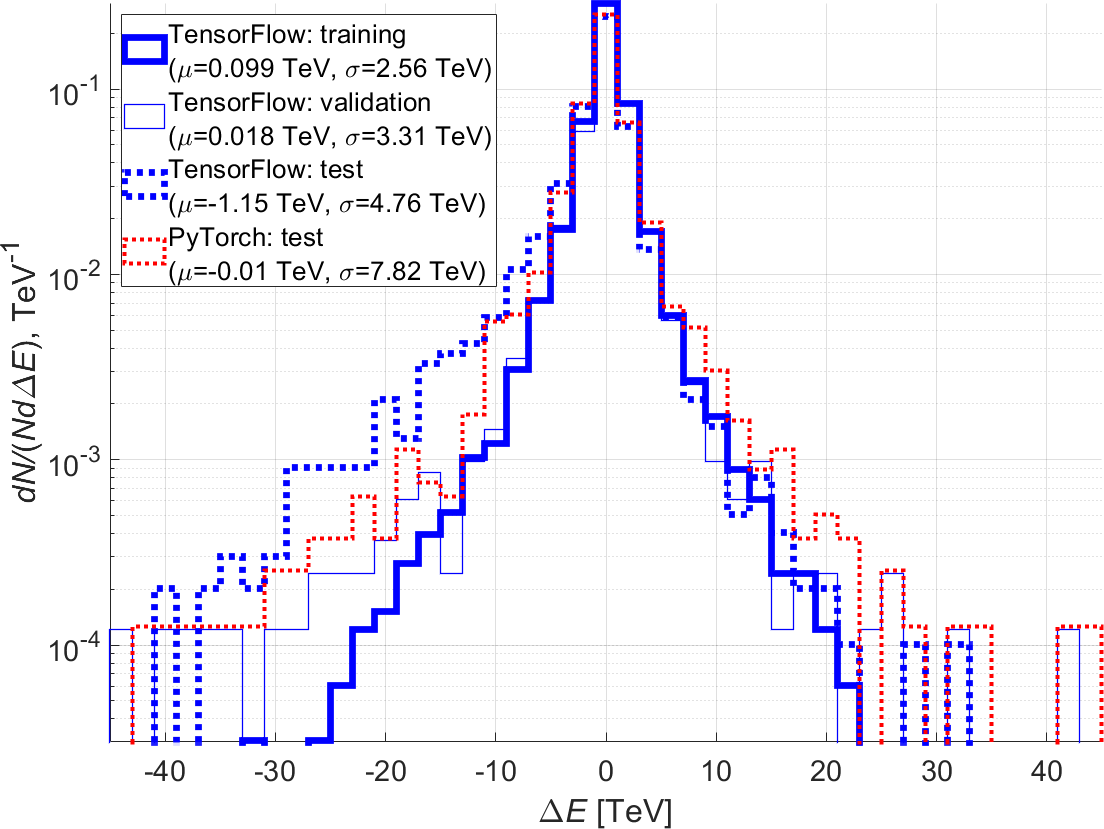}
  %\caption{second}
%  \end{minipage}
%\end{left}
\end{center}
\caption{Absolute energy error distributions.}%: TensorFlow (left), PyTorch (right).} 
\label{figdeltaE}
\end{figure}

\begin{figure}
\begin{center}
 \begin{minipage}{1\textwidth}
  \includegraphics[width=14pc]{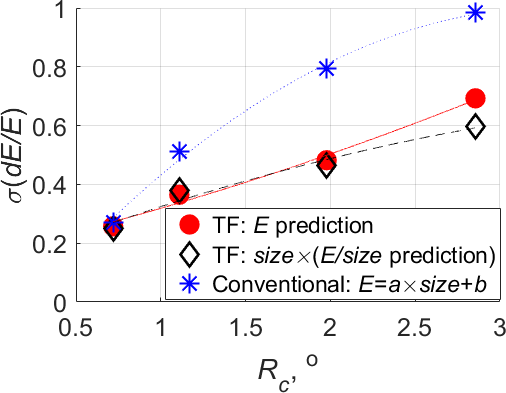}
  \includegraphics[width=14pc]{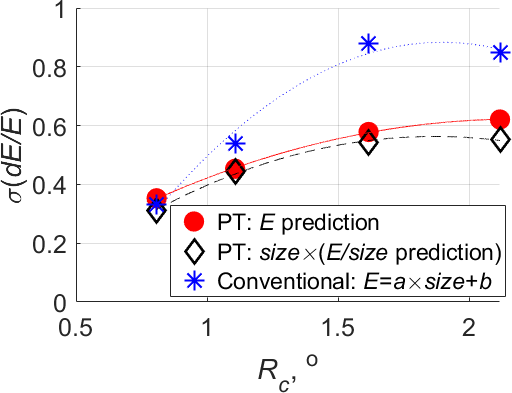}
  %\caption{second}
  \end{minipage}
\end{center}
\caption{Relative energy error vs angular distance of the image: TensorFlow (`TF', left), PyTorch (`PT', right).} 
\label{figE}
\end{figure}

%\begin{figure}
%\includegraphics[width=\textwidth]{dE_vs_Rc_CNNTFcrop.png}
%\caption{Relative energy error vs angular distance of the image.} \label{fig2}
%\end{figure}

\subsection{Energy error}
The predicted energy RMSE is 2.56 TeV for training sample, 3.31 TeV for validation dataset, and 4.76 TeV for the TensorFlow test sample (7.82 TeV for the PyTorch test sample). The dependence of the predicted energy on the primary energy is shown in Figure \ref{reg_im}. Absolute error distribution is presented in Figure \ref{figdeltaE}.

The accuracy of the energy estimate depends on the image distance from the camera center, which corresponds to the distance of the gamma-ray induced shower to the telescope. The energy error is presented in Figure \ref{figE} for various angular distances from the image centre of gravity to the centre of the camera's field of view. This angular distance is strongly correlated with the distance from the gamma-ray-induced shower to the telescope, but is measurable in experiment unlike the unknown distance to the telescope. The angular distance of $\sim$$1.5^o$ corresponds roughly to $\sim$100--150 m \cite{18}. 

%\begin{figure}
%\begin{center}
%% \begin{minipage}{.99\textwidth}
%%  \includegraphics[width=\textwidth]{pred_en_on_true_en}
%  \includegraphics[width=0.8\textwidth]{rms_on_dist}
%  %\caption{second}
%% \end{minipage}
%\end{center}
%\caption{Relative energy error vs angular distance of the image for various CNN stages.} 
%\label{figEin}
%\end{figure}

Though for the nearest gamma-rays there is no improvement over the simplest conventional technique of a linear proportionality to the image size (section \ref{secImplem}), for the distances above $1^o$ CNN gives significantly better results, and especially does the CNN predicting the ratio of the energy to the image size instead of predicting the energy itself. However, it is not the optimal way to incorporate the image size information in the CNN, and therefore the energy estimation still contains some potential to further improve accuracy.

%The energy error for the training and validation datasets in comparison with the test one is plotted in Figure \ref{figEin}.

\section{Conclusion}
Convolutional neural networks were implemented to solve two important tasks of data analysis in gamma-ray astronomy: cosmic ray background suppression and gamma-ray energy estimation. Background rejection quality strongly depends on the learning sample size but in any case is substantially higher than for conventional techniques. The energy estimation achieves significantly better accuracy than conventional approach for gamma-rays incident in the area outside of a narrow circle around the telescope ($\sim$100--150 m on the ground or $\sim$1--1.5$^o$ on the camera plane). Because of the wide acceptance of the TAIGA-IACT camera, this technique is capable of measuring energy of most gamma-rays detected by the installation. 

We also note that there is still considerable potential to further improve the results by taking into account the hexagonal pixel shape and increasing training sample size by one order of magnitude, which is a challenge for the immediate future.  

%\begin{theorem}
%This is a sample theorem. The run-in heading is set in bold, while
%the following text appears in italics. Definitions, lemmas,
%propositions, and corollaries are styled the same way.
%\end{theorem}
%
% the environments 'definition', 'lemma', 'proposition', 'corollary',
% 'remark', and 'example' are defined in the LLNCS documentclass as well.
%
%\begin{proof}
%Proofs, examples, and remarks have the initial word in italics,
%while the following text appears in normal font.
%\end{proof}
%For citations of references, we prefer the use of square brackets
%and consecutive numbers. Citations using labels or the author/year
%convention are also acceptable. The following bibliography provides
%a sample reference list with entries for journal
%articles~\cite{ref_article1}, an LNCS chapter~\cite{ref_lncs1}, a
%book~\cite{ref_book1}, proceedings without editors~\cite{ref_proc1},
%and a homepage~\cite{ref_url1}. Multiple citations are grouped
%\cite{ref_article1,ref_lncs1,ref_book1},
%\cite{ref_article1,ref_book1,ref_proc1,ref_url1}.

\end{document}